\documentclass[a4paper]{jpconf}
\usepackage{graphicx}
\usepackage{subcaption}
\usepackage{hyperref}
\hypersetup{ 
	colorlinks=false,
	linkcolor=black,
	anchorcolor=black,
	citecolor=black,
	filecolor=black,
	menucolor=black,
	runcolor=black,
	urlcolor=black,
	hidelinks=true
}
\begin{document}

\title{\href{https://indico.cern.ch/event/1106990/contributions/4998151/}{Cluster counting algorithms for particle identification at future colliders}}
	
	\author{Brunella D'Anzi$^{1,2}$, Gianluigi Chiarello$^{3}$, Alessandro Corvaglia$^{3}$, Nicola De Filippis$^{2,5}$, Walaa Elmetenawee$^{1,2}$, Francesco De Santis$^{3,4}$, Edoardo Gorini$^{3}$, Francesco Grancagnolo$^{3}$, Marcello Maggi$^{2}$, Alessandro Miccoli$^{3}$, Marco Panareo$^{3}$, Margherita Primavera$^{3}$, Andrea Ventura$^{3,4}$,  Shuiting Xin$^{6,7}$,  Fangyi Guo$^{6,7}$, and Shuaiyi Liu$^{6,7}$ }
\address{$^1$ Dipartimento di Fisica, Università di Bari Aldo Moro, Via E. Orabona n.4, I-70126 Bari,Italy }
\address{$^2$ Istituto Nazionale di Fisica Nucleare Sezione di Bari, Via E. Orabona n.4, I-70126 Bari, Italy}
\address{$^3$ Istituto Nazionale di Fisica Nucleare Sezione di Lecce, Via Arnesano, 73100 Lecce, Italy}
\address{$^4$ Dipartimento di Matematica e Fisica, Università del Salento, Via Arnesano, 73100 Lecce, Italy}
\address{$^5$ Dipartimento di Fisica, Politecnico di Bari, Via Edoardo Orabona, 4, Bari BA, 70126, Italy}
\address{$^6$ Chinese Academy of Sciences, 19A Yuquan Road, Shijing District, Beijing, 100049, China}
\address{$^7$ Institute of High Energy Physics, 19B, Yuquan Road, Shijing District, Beijing, 100049, China}

\ead{\href{mailto:brunella.danzi@ba.infn.it}{brunella.danzi@ba.infn.it}}

\begin{abstract}
Recognition of electron peaks and primary ionization clusters in real data-driven waveform signals is the main goal of research for the usage of the cluster counting technique in particle identification at future colliders. The state-of-the-art open-source algorithms fail in finding the cluster distribution Poisson behaviour even in low-noise conditions. In this work, we present cutting-edge algorithms and their performance to search for electrons peaks and identify ionization clusters in experimental data using the latest available computing tools and physics knowledge. 
\end{abstract}

\section{Introduction}

The large statistical fluctuations in the ionization energy loss high energy physics process by charged particles in gaseous detectors implies that many measurements are needed along the particle track to get a precise mean, and this represent a limit to the particle separation capabilities that should be overcome in the design of future colliders. The cluster counting (CC) technique (dN/dx) represents a valid alternative which takes advantage of the Poisson nature of the primary ionization process and offers a more statistically robust method to infer mass information. 

It consists in singling out, in ever recorded detector signal, the electron peak structures related to the arrival of the electrons belonging to a single primary ionization act (cluster) on the anode wire. Simulation studies by using Garfield++ and Geant4 prove that the cluster counting allows to reach a resolution two times better than traditional dE/dx method over a wide momentum range in the use-case of a helium-based drift chamber  \cite{Simulationarticle}. To validate the simulations results, two beam tests at CERN/H8 facility, using a muon beam ranging from 40 GeV/c to 180 GeV/c on a setup made of 1 cm, 1.5 and 2 cm size drift tubes, equipped with Mo, Al, and W sense wires in the 10–40 $\mu$m diameter range, have been performed by collecting data with the gas mixtures He/iC$_{4}$H$_{10}$ 90/10, 80/20, and 85/15 by making a scan in high voltage, sampling rate and complementary angle between the wire direction and ionizing tracks (15° to 60°) with a total of more than 800k events and a $2 \times 10^5$ gas gain \cite{PisaProceeding,ICHEP}. 

To apply the cluster counting technique successfully, the pulses associated to electrons from different clusters should have a small probability of overlapping in time, but the time distance between electrons coming from the same cluster should be short enough to prevent over-counting. The trade-off for an efficient CC, in helium-based drift chamber case, requires the usage of an high front-end electronics bandwidth and an high sampling rate, respectively of at least 1 GHz and a 1 GSa/s at 12 bits, entirely based on a DRS system \cite{DRS}, which has been crucial in our measurements. Moreover, retrieving the correct electron and clusters time arrival information requires advanced techniques able to spot peaks above the noise level in discretized signals.

In this work, we show the data analysis results concerning the ascertainment of cluster counting technique Poisson nature, the establishment of the most reliable electrons clustering algorithms, and the identification of limiting effects for a fully efficient cluster counting, like the space charge density around the sense wire and the dependency on attachment and recombination effects.
\section{Electron Peak Finding Algorithms}\label{algorithms}
\begin{center}
	\begin{figure}
		\begin{subfigure}{0.5\textwidth}
			\includegraphics[width=\linewidth]{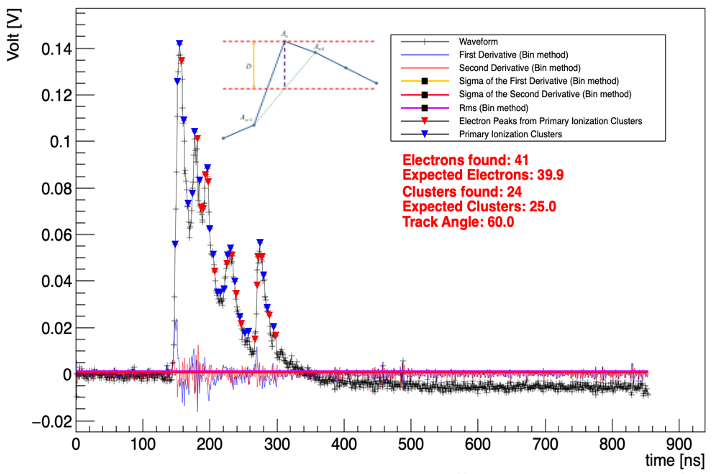}
			\caption{} \label{fig:WaveformDERIV}
		\end{subfigure}\hspace*{\fill}   
		\begin{subfigure}{0.5\textwidth}
			\includegraphics[width=\linewidth]{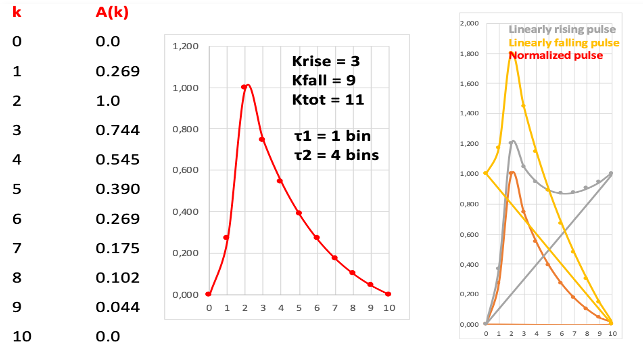}
			\caption{} \label{fig:RTA}
		\end{subfigure}\hspace*{\fill} 
		\caption{\small{(\ref{fig:WaveformDERIV}) Example of a real waveform collected from 1 cm size drift tube (90/10 He/iC$_{4}$H$_{10}$, muon momentum 165 GeV, Track Angle 60°, sampling rate 1.2 GSa/s, sense wire diameter 20 $\mu m$). The blue and red arrows represent the cluster and electron peaks found with the DERIV algorithm. First, second derivative and the associated errors are drawn as indicated from the legenda.The expected number of electrons and clusters is computed by multiplying the number of cluster/cm for a MIP \cite{Clustersizedistributions}, the drift tube size, the inverse of the track angle and a relativistic rise correction factor \cite{EnergyLoss}. (\ref{fig:RTA}): Sketch of a electron template pulse used to scan the waveforms by the RTA algorithm.}}
	\end{figure}
\end{center}
Two electron peak finding methods have been used to retrieve the results shown in \autoref{results}. Before applying them, a preprocessing phase for data collected during beam tests is needed to set the waveform baseline to zero. For this purpose, we define and subtract to each waveform a root mean square (r.m.s.) error performing an average over its first 30 ns in which only noise effects are present. 

The first technique used is a second derivative-based (DERIV) algorithm. It computes the first and second derivative as the ratio between amplitude average over two consecutive bins and two times their bin size depending on the sampling rate. Subsequently, it requires that, at the peak candidate position, the derivatives are less than a scanned quantity proportional to the waveform r.m.s. ($\epsilon$) and they increase (decrease) before (after) the peak candidate position of a $\epsilon$-based small quantity. R.m.s.-based lower limit boundaries on the amplitude at peak candidate positions and on the voltage difference between the peak candidates amplitude and the previous (next) bin ones are imposed as well (see the example in \autoref{fig:WaveformDERIV}).

A second methodology, the running template algorithm (RTA) is under test. It defines an electron pulse template having a raising ($K_{rise}$) and falling ($K_{fall}$) exponential over a fixed number of bins (K$_{tot}$) and digitized (A(k)) to mimic the data behaviour (see the example in \autoref{fig:RTA}). The algorithm scans the waveform and runs over the $K_{tot}$ bins window by comparing the electron pulse shape to the normalized data, builds an agreement assessment quantity ($\chi^{2}$) and defines a cut on it. RTA subtracts the found peak to the signal spectrum and iterates the search until no new peaks are found. The electron pulse is fine-tuned for 1.2 GSa/s setting and its sampling rate dependence is under implementation.

\begin{center}
	\begin{figure}
		\begin{subfigure}{0.45\textwidth}
			\includegraphics[width=\linewidth]{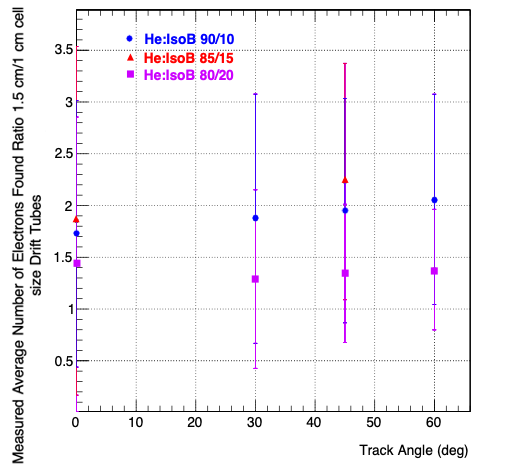}
			\caption{} \label{fig:ElectronRatioPlotAngleScan}
		\end{subfigure}\hspace*{\fill} 
		\begin{subfigure}{0.45\textwidth}
			\includegraphics[width=\linewidth]{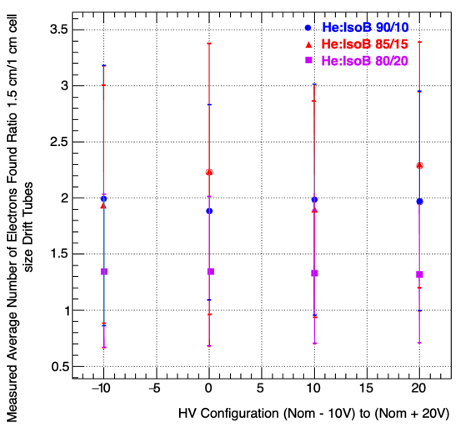}
			\caption{} \label{fig:ElectronRatioPlotHVScan}
		\end{subfigure}\hspace*{\fill}  
		\caption{\small{Ratio plots between the average number of electrons found by the DERIV algorithm for 1 cm and 1.5 cm cell drift tubes as a function of the angle (\ref{fig:ElectronRatioPlotAngleScan}) complementary to the one formed between the sense wire and the muon track passed in the detector and the drift tube high voltage configuration sets for different helium-based gas mixtures (\ref{fig:ElectronRatioPlotHVScan}).}}
	\end{figure}
\end{center}

\begin{center}
	\begin{figure}
		\begin{center}
		\begin{subfigure}{0.5\textwidth}
			\includegraphics[width=\linewidth]{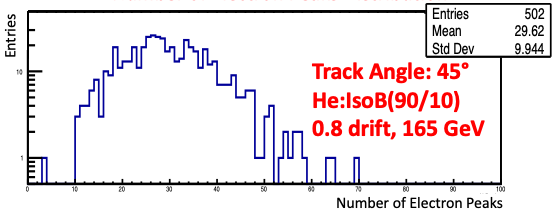}
			\caption{} \label{fig:electrondistribution}
		\end{subfigure}\hspace*{\fill}\\ 
	\end{center}
		\begin{subfigure}{0.54\textwidth}
			\includegraphics[width=\linewidth]{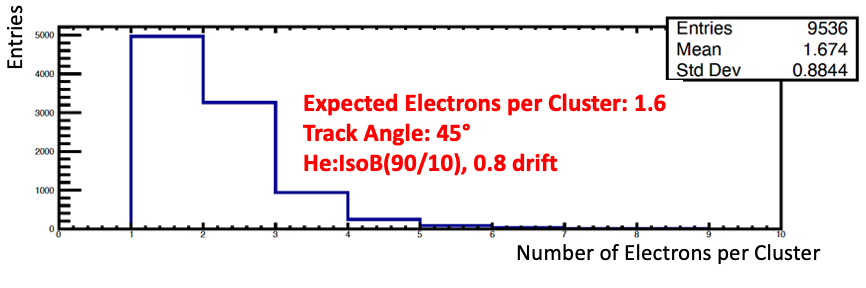}
			\caption{} \label{fig:ElectronsPerCluster}
		\end{subfigure}\hspace*{\fill}
		\hspace*{\fill}\begin{subfigure}{0.5\textwidth}
			\includegraphics[width=\linewidth]{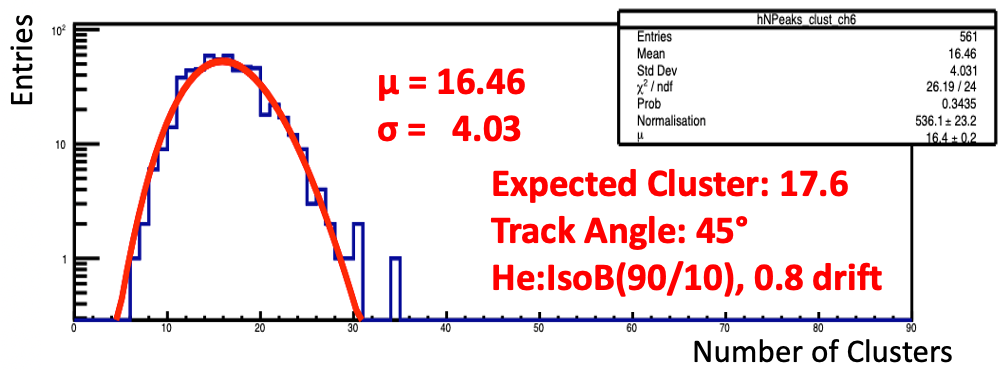}
			\caption{} \label{fig:PoissonClusterdistribution}
		\end{subfigure}\hspace*{\fill}
		\caption{\small{Examples of number of electrons (\ref{fig:electrondistribution}), electron per cluster (\ref{fig:ElectronsPerCluster}) number of clusters (\ref{fig:PoissonClusterdistribution}) distributions found by the DERIV algorithm for 1 cm cell size tubes.}}
	\end{figure}
\end{center}

In both cases, the ratios of measured electron peaks found for different size drift tubes as a function of scan angle, high voltage and  sampling rate are compatible with the expectations within error bars (see \autoref{fig:ElectronRatioPlotAngleScan}, and \autoref{fig:ElectronRatioPlotHVScan}). Those plots are more reliable than the efficiency ones since the expected values for the number of electrons (clusters) strongly depend on assumptions for the number of clusters/cm \cite{Clustersizedistributions} for a minimum ionizing particle (MIP) for a given drift cell-size and the relativistic rise correction factor \cite{EnergyLoss}. Finally, the expected Landau distribution behaviour for the number of electron peaks found has been obtained (see \autoref{fig:electrondistribution}).
\section{Clusterization Strategy and Results}\label{results}
The primary ionization cluster identification (CLUSTER algorithm) consists in associating properly the electron peaks recognized by the algorithms described in \autoref{algorithms}, and defining a time position and amplitude for the cluster. Firstly, it merges the electron peaks being in consecutive bins into a single one to reduce the fake electrons rate.

Secondly, contiguous electron peaks which are compatible with the electrons’ diffusion time are associated to the same ionization cluster. For them, a counter for electrons per each cluster is incremented. The CLUSTER algorithm takes into account that the electrons' diffusion time has a dependence on the electron time arrival  $\approx$ $\sqrt{t_{ElectronPeak}}$ and on the gas mixture used (e.g. the drift velocity changes of $\approx$ 20\% passing from He:IsoB 90/10 to 80/20). 
\begin{center}
	\begin{figure}
		\begin{subfigure}{0.45\textwidth}
			\includegraphics[width=\linewidth]{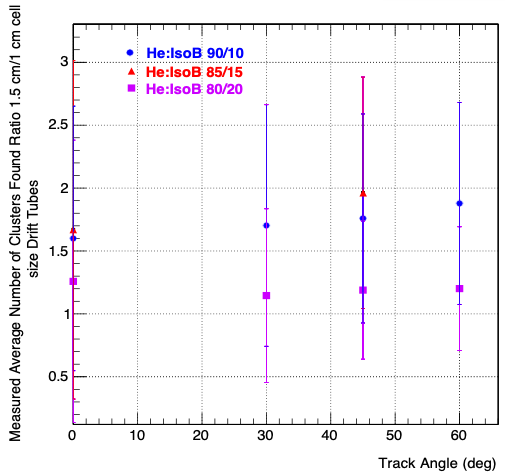}
			\caption{} \label{fig:ClusterRatioPlotAngleScan}
		\end{subfigure}\hspace*{\fill} 
		\begin{subfigure}{0.45\textwidth}
			\includegraphics[width=\linewidth]{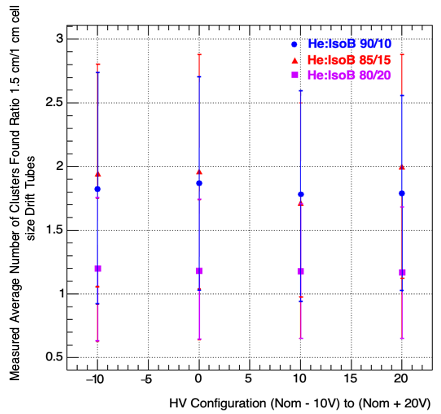}
			\caption{} \label{fig:ClusterRatioPlotHVScan}
		\end{subfigure}\hspace*{\fill}   
		\caption{\small{Ratio plots between the average number of clusters found by the DERIV algorithm for 1 cm and 1.5 cm cell drift tubes as a function of the angle (\ref{fig:ClusterRatioPlotAngleScan}) complementary to the one formed between the sense wire and the muon track passed in the detector and the drift tube high voltage configuration sets for different helium-based gas mixtures (\ref{fig:ClusterRatioPlotHVScan}).}}
	\end{figure}
\end{center}
\begin{center}
	\begin{figure}
		\begin{subfigure}{1.0\textwidth}
			\includegraphics[width=\linewidth]{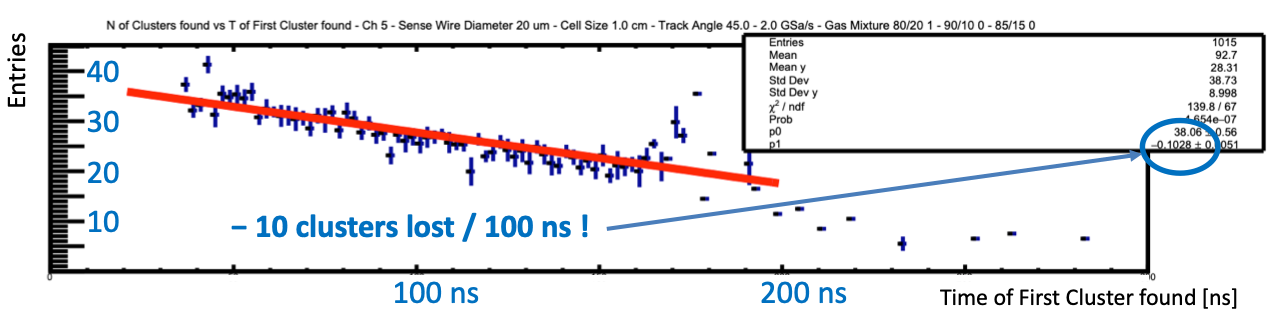}
			\caption{} \label{fig:LossCluster}
		\end{subfigure}\hspace*{\fill}\\ 
		\begin{subfigure}{1.0\textwidth}
			\includegraphics[width=\linewidth]{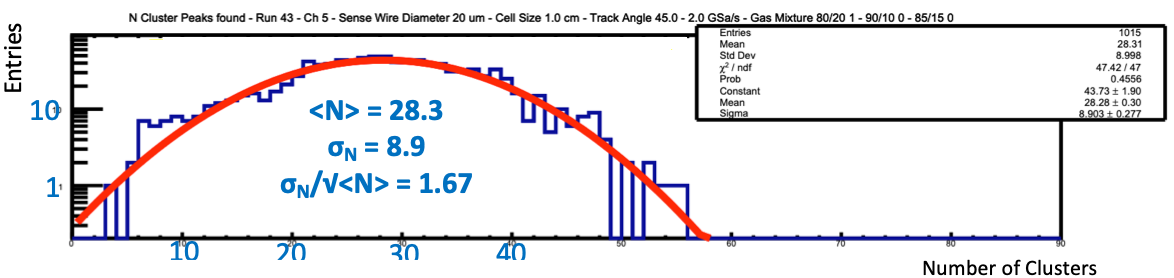}
			\caption{} \label{fig:UncorrectedNumberOfCluster}
		\end{subfigure}\hspace*{\fill} \\  
		\begin{subfigure}{1.0\textwidth}
			\includegraphics[width=\linewidth]{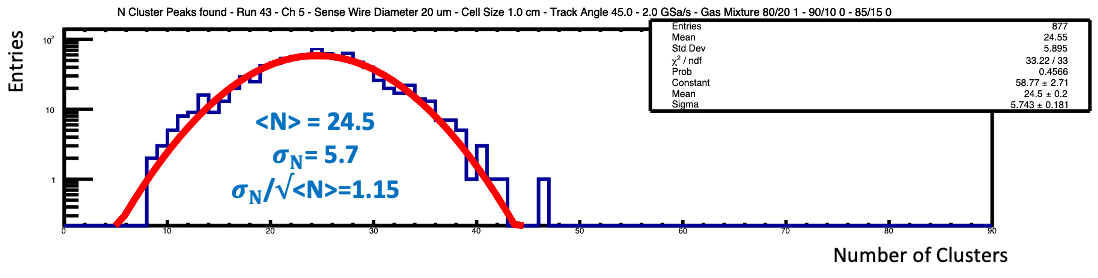}
			\caption{} \label{fig:CorrectedNumberOfCluster}
		\end{subfigure}\hspace*{\fill}   
		\caption{\small{(\ref{fig:LossCluster}): Plot of the average number of clusters (at a certain drift time) found by the DERIV algorithm as a function of the arrival time of the first cluster (drift time-related) for 1 cm cell-size drift tubes. A cluster loss of $\approx$10 clusters every 100 ns has been observed. (\ref{fig:UncorrectedNumberOfCluster}) Example of a no-Poissonian number of cluster distribution due to attachment and recombination effects before the correction implemented in (\ref{fig:CorrectedNumberOfCluster}).}}
	\end{figure}
\end{center}
Finally, the position and amplitude of the clusters corresponds to those of the electrons having the highest amplitude in the cluster. An example of the CLUSTER outcome for the number of electron per clusters distribution and the number of clusters found are shown respectively in \autoref{fig:ElectronsPerCluster} and \autoref{fig:PoissonClusterdistribution}. The ratio plots are in good agreement with expectations as well (see \autoref{fig:ClusterRatioPlotAngleScan}). Moreover, results on the electron drift velocity values can be retrieved from the first cluster drift time distribution and they are compatible with the expected values (e.g. of $\approx 2.5 cm/ \mu s$ for 1 cm (2 cm) cell size tubes, taking into account their maximum impact parameter of 0.5 cm (0.9 cm)). From \autoref{fig:LossCluster} we spotted the combined effect of recombination, electron attachment and electric field suppression due to space charge affect the electron finding peak algorithm performance. Cuts on DERIV and RTA algorithms, which were optimized before including the recombination and attachment effects, need to be reformulated (see an example of uncorrected and corrected cluster distribution in \autoref{fig:UncorrectedNumberOfCluster} and \autoref{fig:CorrectedNumberOfCluster}). Also, these corrections, strongly depend on the drift length and, therefore, on the drift tube size and must be calculated for each different drift tube configuration.

\section{Conclusions and Outlook}
Despite the challenging environment for the search of electron peaks in real-data waveforms, promising cluster counting algorithms have been tested and feature good performance to be used in tracking gaseous detectors prototypes for future colliders. Calibration is under going to take into account attachment and recombination effects in helium-based drift chambers.
\section{Acknowledge}
This project has received funding from the European Union’s Horizon 2020 Research and Innovation programme under Grant Agreement no. 871072.

\smallskip


\medskip
\section*{References}
\begin{thebibliography}{9}
	\bibitem{Simulationarticle}
	F. Cuna \etal Simulation of particle identification with the cluster counting technique Proceeding at LCWS2021 \verb|arXiv:2105.07064 [physics.ins-det]|.
	\bibitem{PisaProceeding}
	C. Caputo \etal Particle identification with the cluster counting technique for the IDEA drift chamber Proceeding Nucl. Instrum. Meth. A \textbf{1048} (2023) 167969 \verb|doi:10.1016/j.nima.2022.167969|.
	\bibitem{ICHEP}
	F. Cuna \etal Particle identification with the cluster counting technique for the IDEA drift chamber PoS ICHEP2022 \textbf{335} (2022)
	\verb|doi:10.22323/1.414.0335|.
	\bibitem{DRS}
	S. Ritt \etal Application of the DRS chip for fast waveform digitizing,''
	Nucl. Instrum. Meth. A \textbf{623} (2010), 486-488
	\verb|doi:10.1016/j.nima.2010.03.045|.
	\bibitem{Clustersizedistributions}
	H. Fischle \etal Experimental determination of ionization cluster size distributions in counting gases Nucl. Instrum. Meth. A \textbf{301} (1991) 202-214 \verb|doi:10.1016/0168-9002(91)90460-8|.
	\bibitem{EnergyLoss}
	R. G. Kepler \etal Relativistic increase of energy loss by ionization in gases. Nuovo Cim \textbf{7} 71–86 (1958) \verb|https://doi.org/10.1007/BF02746883|.
	
\end{thebibliography}
\end{document}